\documentclass[showpacs,preprint,amsmath,amssymb]{revtex4}

\begin{document}
\title{Isoscalar giant monopole resonance and its overtone in microscopic and
macroscopic models}
\author{S. Shlomo$^{1)}$, V. M. Kolomietz$^{1,2)}$ and B.K. Agrawal$^{1)}$}
\address{$^{1)}$Cyclotron Institute, Texas A\&M University,
College Station, Texas 77843, USA\\
$^{2)}$Institute for Nuclear Research, Kiev 03680, Ukraine}

\begin{abstract}
We calculate the transition density for the overtone of the isoscalar giant
monopole resonance (ISGMR) from the response to  an appropriate external
field $\sim \hat{f}_\xi({\bf r})$ obtained using the semiclassical
fluid dynamic approximation  and the Hartree-Fock (HF) based
random phase approximation (RPA).  We determine the  mixing parameter
$\xi$ by maximizing  the ratio of the  energy-weighted sum for the overtone mode  to the
total energy-weighted sum rule and derive a simple expression for the
macroscopic transition density associated with the overtone mode.  This  
macroscopic transition density  agrees well  with that obtained from
the HF-RPA calculations.  We also point out that the ISGMR and its overtone can be
clearly identified by considering the response to the electromagnetic
external field $\sim j_0(qr)$.

\end{abstract}
\pacs{PACS numbers : 21.60Jz, 24.30.Cz, 26.60.Ev, 24.10.Nz}
\maketitle


\bigskip \bigskip

\section{Introduction}

\bigskip

The properties of a giant resonance in nuclei are commonly determined from
the  distorted wave Born approximation (DWBA) analysis of its excitation
cross-section by inelastic scattering of a certain projectile.  The
transition potential required in actual implementation of DWBA calculation
is usually obtained by  convoluting  the projectile-nucleus interaction
with the transition density associated with the giant resonance. The
relevant transition density can be obtained from a microscopic theory of
the giant resonance, such as the Hartree-Fock (HF) based random phase
approximation (RPA).  However, the use of a macroscopic transition
density $\rho _{{\rm tr}}^{{\rm macr}}({\bf r})$ greatly simplifies the
application of the giant multipole resonance theory to the analysis of
the experimental data.  The simple form of the transition density
\begin{equation}
\rho _{{\rm tr}}^{{\rm sc}}({\bf r})=\alpha _{0}\left( 3+r\frac{d}{dr}%
\right) \rho _{{\rm eq}}(r)Y_{00}({\bf \hat{r}}), \label{tr7}
\end{equation}
obtained by the scaling approximation,
is a well-known  example of the macroscopic transition density $\rho _{{\rm tr}}^{{\rm %
macr}}({\bf r})$ commonly used in the case of the isoscalar giant monopole
resonance (ISGMR) \cite{blaizot}. The transition density of  Eq. (\ref{tr7}) 
nicely agrees with the ISGMR transition density obtained in 
microscopic HF-RPA calculations. It has a one-node structure,  satisfying
the condition of particle number conservation $\int \rho _{{\rm tr}}^{{\rm
sc}}({\bf r})\ d{\bf r}=0.$
Unfortunately the  scaling consideration can not be extended to the
overtone of the ISGMR,  where $\rho _{{\rm tr}}^{{\rm macr}}({\bf r})$ has a 
two-node structure.  To derive the macroscopic transition density 
$\rho _{{\rm tr}}^{{\rm macr}}({\bf r})$ in this more general case, one can
use the well-known method \cite{bm,rs} of determining $\rho _{{\rm tr}%
}^{{\rm macr}}({\bf r})$ from the local sum rule which is exhausted by one
collective state with the appropriate choice of the transition operator $%
\hat{f}_{\xi }({\bf r})$. However,  in the quantum random phase approximation,
the highly excited collective modes are strongly fragmented over 
 a wide range of energy  and a special averaging procedure must be employed to
determine the macroscopic transition density corresponding to  an average  collective
excitation. In this respect, the semiclassical Fermi-liquid approach (FLA) 
\cite{kosh} is more appropriate. Both the main ISGMR and its overtone are
well-defined within the FLA as  single resonance states. This fact
enables us to derive the transition operator $\hat{f}_{\xi }({\bf r})$\
simply by maximizing the fraction of the energy-weighted sum rule (FEWSR)  exhausted
by the single overtone.

In the  present work  we suggest a procedure   to derive the macroscopic
transition density for the ISGMR overtone using both the
HF based RPA and the Fermi-liquid approaches.
We remark that some  preliminary results of this investigation were
presented in Ref. \cite{Shlomo01} (see also Ref. \cite{Gorelik02}).

\bigskip

\section{Quantum derivations}

\bigskip

{\bf \ }The transition density $\rho _{{\rm tr,n}}({\bf r})$ for a certain
eigenstate $|n\rangle $ of a nucleus with  $A$\ nucleons is given by \ \ ~ 
\begin{equation}
\rho _{{\rm tr,n}}({\bf r})=\langle 0|\hat{\rho}({\bf r})|n\rangle ,
\label{tr0}
\end{equation}
where $\hat{\rho}({\bf r})=\sum_{i=1}^{A}\delta ({\bf r-r}_{i})$ is the
particle density operator and $|0\rangle $\ represents the  ground state of the nucleus.
The transition density reflects the internal structure of the nucleus and
does not depend on the external field. However, a problem arises if one
intends to derive the transition density $\rho _{{\rm tr}}({\bf r})$ for a
group of the\ thin-structure resonances in the giant multipole resonance
(GMR) region. An appropriate averaging procedure is necessary in this case
and $\rho _{{\rm tr}}({\bf r})$\ can be evaluated if the nucleus is placed
in an external field 
\begin{equation}
V_{{\rm ext}}\sim \ \hat{F}(\{{\bf r}%
_{i}\})=\sum_{i=1}^{A}\hat{f}({\bf r}_{i}),
\end{equation}
 where the transition operator $%
\hat{F}(\{{\bf r}_{i}\})$ is so chosen  that it  provides a preferable excitation
of the above mentioned thin-structure resonances. \ 

Let us introduce the local strength function

\begin{equation}
S({\bf r},E)=\sum_{n}\ \langle 0|\hat{\rho}({\bf r})|n\rangle \ \langle n|%
\hat{F}|0\rangle \ \delta (E-E_{n})  \label{sre}
\end{equation}
and the energy smeared local strength function $\widetilde{S}({\bf r},E)$
defined near the GMR energy $E_{R}$\ by 
\begin{equation}
\widetilde{S}({\bf r},E_{R})=\frac{1}{\Delta E}\int_{E_{R}-\Delta
E/2}^{E_{R}+\Delta E/2}dE\ S({\bf r},E).  \label{sre1}
\end{equation}
The corresponding strength functions are given by 
\begin{equation}
S(E)=\int d{\bf r}\ \hat{f}({\bf r})\ S({\bf r},E)=\sum_{n}\ \left| \int d%
{\bf r}\ \hat{f}({\bf r})\ \rho _{{\rm tr,n}}({\bf r})\right| ^{2}\ \delta
(E-E_{n}),\qquad  \label{se0}
\end{equation}
and 
\begin{equation}
\widetilde{S}(E)=\int d{\bf r}\ \hat{f}({\bf r})\ \widetilde{S}({\bf r},E).
\label{se0a}
\end{equation}
Let us assume, for the moment, that the operator $\hat{F}(\{{\bf r}_{i}\})$\
excites only a single state $|D\rangle $, within the energy interval $%
E_{R}\pm \Delta E/2$. The corresponding transition density $\rho _{{\rm tr,D}%
}({\bf r})=\langle 0|\hat{\rho}({\bf r})|D\rangle $ is then given by the
following {\it exact} expression 
\begin{equation}
\rho _{{\rm tr,D}}({\bf r})=\frac{\Delta E}{\sqrt{\widetilde{S}(E_{R})\
\Delta E}}\widetilde{S}({\bf r},E_{R}).  \label{tr1}
\end{equation}
We will extend expression (\ref{tr1}) to the case of a group of
the thin-structure resonances in the GMR region which are excited by the
operator $\hat{F}(\{{\bf r}_{i}\})$ and define the smeared transition
density $\widetilde{\rho }_{{\rm tr,R}}({\bf r})$\ as 
\begin{equation}
\widetilde{\rho }_{{\rm tr,R}}({\bf r})=\frac{\Delta E}{\sqrt{\widetilde{S}%
(E_{R})\ \Delta E}}\widetilde{S}({\bf r},E_{R}).  \label{tr2}
\end{equation}
Note that  Eq. (\ref{tr2}) is associated with the strength in the region of $%
E_{R}\pm \Delta E/2$ and is consistent with the smeared strength function $%
\widetilde{S}(E_{R})$\ for a single resonance state. That is  (see also Eq. (%
\ref{se0})), \ 
\begin{equation}
\widetilde{S}(E_{R})=\frac{1}{\Delta E}\left| \int d{\bf r}\ \hat{f}({\bf r}%
)\ \widetilde{\rho }_{{\rm tr,R}}({\bf r})\right| ^{2}.  \label{se1}
\end{equation}

We\ also point out that with the Lorentz's function 
\begin{equation}
g_{\gamma }(E,E_{R})=\frac{1}{\pi }\frac{\gamma }{(E-E_{R})^{2}+\gamma ^{2}},
\label{lorentz}
\end{equation}
the energy smeared $\widetilde{S}({\bf r},E_{R})$ is given by 
\begin{equation}
\widetilde{S}({\bf r},E_{R})=\int_{-\infty }^{\infty }\ dE\ S({\bf r},E)\
g_{\gamma }(E,E_{R}),  \label{s2}
\end{equation}
and the smeared transition density $\widetilde{\rho }_{{\rm tr,R}}({\bf r})$ is
obtained from 
\begin{equation}
\widetilde{\rho }_{{\rm tr,R}}({\bf r})=\frac{\pi \gamma }{\sqrt{\widetilde{S%
}(E_{R})\ \pi \gamma }}\widetilde{S}({\bf r},E_{R}).  \label{tr3}
\end{equation}
The consistency condition, Eq.  (\ref{se1}), then reads 
\begin{equation}
\widetilde{S}(E_{R})=\frac{1}{\pi \gamma }\left| \int d{\bf r}\ \hat{f}({\bf %
r})\ \widetilde{\rho }_{{\rm tr,R}}({\bf r})\right| ^{2}.  \label{se3}
\end{equation}
\bigskip

In the quantum RPA, the local strength function $S({\bf r},E)$ is related to
the RPA Green's function $G({\bf r}\,^{\prime },{\bf r},E)$ by \cite
{bets,shbe} 
\begin{equation}
S({\bf r},E)=\int \hat{f}({\bf r}\,^{\prime })\left[ {\frac{1}{\pi }}%
\mbox{Im}G({\bf r}\,^{\prime },{\bf r},E)\right] \,d{\bf r}\,^{\prime }\ .
\label{se4}
\end{equation}
For the isoscalar monopole and dipole excitations, the transition operator $%
\hat{f}({\bf r})$ is taken in the form of 
\begin{equation}
\label{fl0}
\hat{f}({\bf r})\equiv \hat{f}_{\xi }({\bf r})=f_{\xi }(r)Y_{00}({\bf 
\hat{r}})\quad {\rm for\quad }L=0,
\end{equation}
{\rm and}
\begin{equation}
\hat{f}({\bf r}%
)\equiv \hat{f}_{\eta }({\bf r})=f_{\eta }(r)Y_{10}({\bf \hat{r}}%
)\quad {\rm for}\quad L=1,  \label{f1}
\end{equation}
with an appropriate choice of the radial functions $f_{\xi }(r)$ and $%
f_{\eta }(r)$, see below. In the following,  the quantum transition density for
the main ISGMR and its overtone is evaluated using the  Eq. (\ref{tr2}%
) with $E_{R}\pm \Delta E/2$\ taken separately for the ISGMR and the overtone
regions.

\section{Macroscopic transition density}

\bigskip Let us consider the local energy-weighted sum $M_{1}({\bf r})$
given by (see Eq. (\ref{sre})) 
\begin{equation}
M_{1}({\bf r})=\int_{0}^{\infty }dE\ ES({\bf r},E)=\sum_{n}\ E_{n}\ \langle
0|\hat{\rho}({\bf r})|n\rangle \ \langle n|\hat{F}|0\rangle ,\qquad
(E_{0}=0).  \label{m1r1}
\end{equation}
The continuity equation provides the following sum rule \cite{rs} 
\begin{equation}
M_{1}({\bf r})=-\frac{\hbar ^{2}}{2m}{\bbox{\nabla}}\left( \rho _{{\rm eq}}{%
\bbox{\nabla}}\hat{f}({\bf r})\right) .  \label{m1r2}
\end{equation}
Let us assume that only one state $|D\rangle $\ exhausts the sum rule Eq.  (\ref
{m1r2}). Then for the corresponding (''macroscopic'') transition density, $%
\rho _{{\rm tr}}^{{\rm macr}}({\bf r})$, we have from Eqs. (\ref{m1r1}) and (%
\ref{m1r2}) the following expression 
\begin{equation}
\rho _{{\rm tr}}^{{\rm macr}}({\bf r})=\ \langle 0|\hat{\rho}({\bf r}%
)|D\rangle =\alpha _{0}\ {\bbox{\nabla}}\left( \rho _{{\rm eq}}{\bbox{\nabla}%
}\hat{f}({\bf r})\right) ,  \label{tr5}
\end{equation}
where the normalization coefficient $\alpha _{0}$\ can be found using the
energy-weighted sum rule (EWSR) 
\begin{equation}
m_{1}=\int d{\bf r}\ \hat{f}({\bf r})\ M_{1}({\bf r})=\frac{\hbar ^{2}}{2m}%
\int d{\bf r\ }\rho _{{\rm eq}}\left( {\bbox{\nabla}}\hat{f}({\bf r})\right)
^{2}.  \label{m1}
\end{equation}
Taking into account that 
\[
\langle D|\hat{F}|0\rangle =\left( \int d{\bf r\ }\hat{f}({\bf r})\rho _{%
{\rm tr}}^{{\rm macr}}({\bf r})\right) ^{\ast }, 
\]
we obtain 
\begin{equation}
|\alpha _{0}|=(\hbar ^{2}/2m)(E_{D}m_{1})^{-1/2}.  \label{beta}
\end{equation}
Thus, the macroscopic transition density $\rho _{{\rm tr}}^{{\rm macr}}({\bf %
r})$ of Eq. (\ref{tr5})  coincides with the quantum transition
density for a certain state $|D\rangle $\ if  the  single state $|D\rangle $\
exhausts all the EWSR, Eq. (\ref{m1}), associated with the transition operator $%
\hat{f}({\bf r})$. In the case of the transition operator $\hat{f}({\bf r})$%
\ from Eqs. (\ref{fl0}) and  (\ref{f1}),\ the EWSR is given by \cite{bm} 
\begin{equation}
m_{1}=\frac{\hbar ^{2}}{2m}\frac{A}{4\pi }(2L+1)\left[ \left\langle 0\left|
\left( \frac{df}{dr}\right) ^{2}+L(L+1)\left( \frac{f}{r}\right) ^{2}\right|
0\right\rangle \right] .  \label{m1l}
\end{equation}

Assuming that the energy-weighted transition strength (EWTS) $E_{D}\
|\langle D|\hat{F}|0\rangle |^{2}$ for the   resonance at $E_D$ 
fully exhausts the EWSR associated with $\hat{f}({\bf r})=f(r)Y_{LM}({\bf \hat{r}})$%
, we obtain the expression for the macroscopic transition density for the state at $E_D$
from Eq. (\ref{tr5}) \cite{agsh}  
\begin{equation}
\rho _{{\rm tr}}^{{\rm macr}}({\bf r})=-\frac{\hbar ^{2}}{2m}\sqrt{\frac{2L+1%
}{m_{1}E_{D}}}\left[ \frac{1}{r}\frac{d^{2}}{dr^{2}}(r\ f)-\frac{L(L+1)}{%
r^{2}}f+\frac{df}{dr}\frac{d}{dr}\right] \rho _{{\rm eq}}(r)Y_{LM}({\bf \hat{%
r}}).  \label{trl}
\end{equation}
For $L=0$\ we have from Eq. (\ref{trl}) the commonly used ISGMR result of
Eq. (\ref{tr7}). The main (lowest) ISGDR is the spurious state with the
eigenenergy $E_{1^{-}}=0$. The next ISGDR is the overtone. Assuming that the
EWTS for the $1^{-}$\ overtone equals the EWSR associated with the operator 
\begin{equation}
\hat{f}_{\eta }({\bf r}) = (r^3-\eta r)Y_{10}(\hat{\bf r}),
\end{equation}
 we obtain from Eq. (\ref{tr5}) the
macroscopic transition density as 
\begin{equation}
\rho _{{\rm tr,overtone}}^{{\rm macr}}({\bf r})=\alpha _{0}\sqrt{3}\left(
10\ r+(3\ r^{2}-\eta )\frac{d}{dr}\right) \rho _{{\rm eq}}(r)Y_{10}({\bf 
\hat{r}})\qquad {\rm for\qquad }L=1.  \label{tr8}
\end{equation}
For the isoscalar dipole mode, the translation invariance condition is used
for the derivation of $\eta $. This condition\ implies that the center of  mass
of the system can not be affected by internal excitation. We thus have, 
\begin{equation}
\int d{\bf r}\ r\ Y_{10}^*(\hat{\bf r})   \rho _{{\rm tr}}^{{\rm macr}}({\bf r})=0\ \qquad {\rm %
for\qquad }L=1.  \label{cm1}
\end{equation}
From Eqs. (\ref{tr8}) and (\ref{cm1}) one obtains, see also Refs. \cite
{hsz1,h1,d1}, 
\begin{equation}
\eta =\frac{5}{3}\left\langle r^{2}\right\rangle,  \label{eta2}
\end{equation}
where $\left\langle r^{2}\right\rangle =\int_{0}^{\infty }dr\ r^{4}\rho _{%
{\rm eq}}(r)/\int_{0}^{\infty }dr\ r^{2}\rho _{{\rm eq}}(r)$\ is the 
mean square radius.

The  free (mixing) parameter appearing in the transition operator $\hat{f}_\xi({\bf r})$ (similar to 
$\eta$ of Eq. (\ref{f1}) for the $L=1$\ case)  can be determined by 
an appropriate
condition  leading to  a
general method for the evaluation of the transition density for the overtone
mode.  In this work we present this method for the case of the monopole mode, $L=0.$
Let us introduce the transition operator $\hat{f}_{\xi }({\bf r})$\ as 
\begin{equation}
\hat{f}_{\xi }({\bf r})=(r^{4}-\xi r^{2})Y_{00}({\bf \hat{r}})\quad {\rm %
for\quad }L=0.  \label{f4}
\end{equation}
The corresponding\ macroscopic transition density $\rho _{{\rm tr}}^{{\rm %
macr}}({\bf r})$\ is obtained from Eqs. (\ref{tr5}) and (\ref{f4}) as 
\begin{equation}
\rho _{{\rm tr}}^{{\rm macr}}({\bf r})=2\alpha _{0}\left[ 10r^{2}-3\xi
+r\left( 2r^{2}-\xi \right) \frac{d}{dr}\right] \rho _{{\rm eq}}(r)Y_{00}(%
{\bf \hat{r}})\qquad {\rm for\qquad }L=0\quad ({\rm overtone}).  \label{tr9}
\end{equation}
The determination of the parameter $\xi $\ in Eq. (\ref{tr9}) requires an
additional consideration since for the $L=0$ case we have no fundamental
condition such as Eq. (\ref{cm1}) for the  $L=1$\ mode. 
We note, however,  that 
 if we assume that the ISGMR has the transition density of Eq. (\ref{tr7}) and require that 
\begin{equation}
\int d{\bf r} \hat{f}_{\xi}({\bf r})\rho_{\rm tr}^{\rm sc}({\bf r})=0,
\end{equation}
i.e.,  the ISGMR is not excited by the scattering operator 
of Eq. (\ref{f4})  we have 
\begin{equation}
\xi = 2\langle r^4\rangle/\langle r^2\rangle .
\label{xi02}
\end{equation}
Similar result is obtained by imposing the condition  
that the scattering operator $r^2Y_{00}(\hat{\bf r})$ does not excite the 
overtone of the  ISGMR, assuming the transition density of Eq. (\ref{tr9}).

Following the general
requirement for the proper use of Eq. (\ref{tr5}) in the derivation of the
macroscopic transition density $\rho _{{\rm tr}}^{{\rm macr}}({\bf r})$, we
can determine the parameter $\xi $ from the condition that the transition
operator $\hat{f}_{\xi }({\bf r})$\ provides for the single overtone the
maximum fraction of the energy-weighted sum rule  $m_{1}$\ of Eq. (\ref{m1}).

\bigskip

\section{Semiclassical Fermi-liquid approach}

\bigskip

The transition density\ $\rho _{{\rm tr}}({\bf r})$\ and\ the strength
function\ $S(E)$\ can also be evaluated within the semiclassical
Fermi-liquid approach. For a given multipolarity $L$\ and overtone $n,$\ the
FLA transition density is given by \cite{kosh}\ 
\[
\rho _{{\rm tr,}Ln}^{{\rm FLA}}({\bf r})=\alpha _{Ln}\left[ \theta
(R_{0}-r)j_{L}(q_{Ln}r)+\right. 
\]
\begin{equation}
\left. \frac{1-a\ \delta _{L1}}{q_{Ln}}\delta (R_{0}-r)j_{L}^{^{\prime
}}(q_{Ln}R_{0})\right] \rho _{0}Y_{L0}({\bf \hat{r}}),  \label{tr4}
\end{equation}
where $\rho _{0}$ is the bulk density,  $R_{0}$ is the equilibrium nuclear
radius and the  parameter $a$ is determined by the translation invariance
condition (Eq.  (\ref{cm1}))\ in the case of the isoscalar dipole compression mode
and is given by 
\begin{equation}
a=j_{1}(x)/xj_{1}^{^{\prime }}(x),\ \ x=q_{Ln}R_{0}.  \label{a}
\end{equation}
The wave numbers $q_{Ln}$ are derived from the boundary conditions of the
FLA model: the normal component of the tensor pressure\ $\delta P_{\nu \mu }$%
,\ created by a sound wave, on the free surface of the nucleus should be
equal to zero \ 
\begin{equation}
\ \left. \delta P_{rr}\right| _{R=R_{0}}=0.  \label{b1}
\end{equation}
Note that for the case of compression sound modes, the contribution from the
surface tension pressure is negligible and it was omitted in Eq. (\ref{b1}%
). The boundary condition (Eq.  (\ref{b1})) leads to the following secular equation
(see Ref. \cite{koko99}) 
\begin{equation}
\lbrack qr\ j_{0}(qr)-D_{\mu }\,j_{1}(qr)]_{r=R_{0}}=0,\quad \,D_{\mu }={%
\frac{4\,\mu }{m\,\rho _{0}c_{0}^{2}},}\qquad  \label{sec2}
\end{equation}
where  $c_0$ is the zero sound velocity and 
 the coefficient $\mu $\ determines  the contribution from the dynamical
Fermi surface distortion associated with the collective motion in a Fermi
liquid. In the case of a quadrupole distortion of the Fermi surface, one
has \cite{kidks} 
\begin{equation}
\mu ={\rm 
\mathop{\rm Im}%
}{\ (}\frac{\omega \tau }{1-i\omega \tau })P_{{\rm eq}}.  \label{mu}
\end{equation}
Here, $P_{{\rm eq}}\approx (2/5)\epsilon _{F}\rho _{0}$\ \ is the
equilibrium pressure of a Fermi gas, $\epsilon _{F}$\ is the Fermi energy,\
and $\omega $\ and$\ \tau $ are the eigenfrequency and the relaxation time
for sound excitations in the Fermi liquid, respectively. The relaxation time 
$\tau $ is assumed to be frequency dependent because of the memory effect in
the collision integral \cite{abkh}. Following Refs. \cite
{kidks,kolmagpl,koplsh} we take 
\begin{equation}
\tau =4\pi ^{2}\beta \hbar /(\hbar \omega )^{2},  \label{tau1}
\end{equation}
where $\beta $ is the constant related to the differential cross section for
the scattering of two nucleons in the nuclear interior. In the case of
isoscalar sound mode, we will adopt $\beta =1.5$\ {\rm MeV} \cite{k}.\ The
eigenfrequency $\omega $ is obtained from the dispersion relation 
\begin{equation}
{\omega }^{2}-c_{0}^{2}q^{2}+i{\omega \gamma }q^{2}=0,  \label{disp}
\end{equation}
\ where $c_{0}$ is given by 

\begin{equation}
c_{0}^{2}=\frac{1}{9m}(K+12\mu /\rho _{0}).  \label{c0}
\end{equation}
 Here $K$\ is the nuclear incompressibility coefficient and $\gamma $ is the friction
coefficient 
\begin{equation}
\gamma =\frac{4\nu }{3\rho _{0}m},\quad \nu ={\rm 
\mathop{\rm Re}%
}\left( \frac{\tau }{1-i\omega \tau }\right) P_{{\rm eq}}.  \label{fric}
\end{equation}

The  smeared FLA strength function $\widetilde{S}^{{\rm FLA}}(E)$ can be obtained in a way
similar to $\widetilde{S}(E)$, obtained within the quantum approach of Eq. (%
\ref{se3}). That is, 
\begin{equation}
\widetilde{S}^{{\rm FLA}}(E)=\sum_{n}\left| \int \!\!d{\bf r}\,\rho _{{\rm tr,}Ln}^{{\rm %
FLA}}({\bf r})\hat{f}({\bf r})\right| ^{2}g_{\gamma }(E,E_{Ln}).
\label{sfla}
\end{equation}
The smearing function $g_{\gamma }(E,E_{Ln})$ in Eq. (\ref{sfla}) is given
by ($\gamma _{Ln}\ll E_{Ln}$) 
\begin{equation}
g_{\gamma }(E,E_{Ln})=\frac{1}{\pi }\frac{\gamma _{Ln}}{(E-E_{Ln})^{2}+%
\gamma _{Ln}^{2}}\ ,\quad \gamma _{Ln}=\frac{1}{2}\frac{\gamma }{\hbar \
c_{0}^{2}}E_{Ln}^{2}\ .  \label{g1}
\end{equation}
Here, $\gamma _{Ln}$ is the damping parameter due to the viscosity of the
Fermi liquid \ and $E_{Ln}=\hbar 
\mathop{\rm Re}%
(\omega _{Ln})$, where the eigenfrequency $\omega _{n}$\ is obtained\ as a
solution to both the dispersion equation (\ref{disp})\ and the secular
equation (\ref{sec2}). We point out that the amplitude $\alpha _{Ln}\ $in Eq. (%
\ref{tr4})$\ $for the FLA transition density $\ \rho _{{\rm tr,}Ln}^{{\rm FLA%
}}({\bf r})$ is derived as the amplitude of the quantum oscillations 
\begin{equation}
\alpha _{Ln}=\sqrt{\hbar /2B_{L}(q)\ \omega (q)},  \label{a1}
\end{equation}
where $q=q_{Ln}$\ is determined by Eq. (\ref{sec2}) and $B_{L}(q)$\ is the
corresponding mass coefficient with respect to the density oscillations. The
collective mass coefficient $B_{L}(q)$ can be found from the 
\ collective kinetic energy $E_{{\rm kin}}$\ for the particle density
oscillations. The collective kinetic energy is derived as 
\begin{equation}
E_{{\rm kin}}=\frac{1}{2}m\rho _{0}\int d{\bf r\ }v^{2}=\frac{1}{2}B_{L}%
\stackrel{\cdot }{\alpha }_{L}^{2}.  \label{kin1}
\end{equation}
For the compression modes $L=0$, the mass coefficient $B_{0}$\ is given by 
\cite{kosh} 
\begin{equation}
B_{0}\equiv B_{0}(q)=(1/2)m\,\rho _{0}R_{0}^{5}\,x_{0}^{-4}\,\left[
1-j_{0}^{2}(x)-D_{\mu }\ j_{1}^{2}(x)\right] _{x=x_{0}},\qquad {\rm %
for\qquad }L=0,  \label{b10}
\end{equation}
where $x_{0}=qR_{0}$.\ 

\bigskip

\section{Results and Discussions}

\bigskip

We have carried out calculations for the  ISGMR
in the frameworks of HF based RPA  and the semiclassical Fermi
liquid approach as briefly outlined in the preceding sections.
We have evaluated the smeared  {\rm %
FLA} strength function $\widetilde{S}^{{\rm FLA}}(E)$ of Eq. (\ref{sfla}) and the
HF-RPA smeared strength function  $\widetilde{S}(E)$ of Eq. (\ref{se3}) for the 
{\rm ISGMR} in several nuclei, for $\hat{f}_{\xi}({\bf r})$\ from Eq.
(\ref{f4}). 
In the subsequent discussions, the HF-RPA results presented are obtained using the Skyrme
force SkM$^*$ \cite{Bartel82}.
For the RPA  calculations to be highly accurate we discretized the continuum in a large box of size 90 fm
and use a smearing parameter $\gamma=\Gamma/2 = 1.0 $ MeV, in evaluating the RPA Green's function (see Eq. (\ref{se4})
), and allow  particle-hole excitations up to 500
MeV (see Ref. \cite{agsh} for the details).
In case of  the FLA
calculations we have adopted the values  of  $\rho _{0}=0.14$ fm$^{-3}$, $\epsilon _{F}=32.85\ {\rm MeV}$
and \ $R_{0}=1.2\cdot \ A^{1/3}$ fm.
In Table  \ref{fla-rpa} we compare the FLA and RPA results for the
centroid energies $E_{01}$ and $E_{02}$ corresponding to the main and
overtone mode of the ISGMR, respectively. We see from this table that the
FLA and RPA results are in qualitative agreement. The small differences
($< 10\%$) can be understood  by the fact that the centroid energy mainly
depends on the size of the system.
The values of the mean square radii and the higher moments of the
ground state density distribution are smaller in the FLA than the ones
obtained from HF calculations.  
Note also that the ratio $E_{02}/E_{01}$ in both  models considered is 
greater than two ($\sim 2.2 - 2.4$). We point out that $E_{02} > 2 E_{01}$ is due to the Fermi-surface distortion effect as noted earlier  in Ref. \cite{koko99}.

We have performed a comparison of the macroscopic transition density $\rho
_{{\rm tr}}^{{\rm macr}}({\bf r})$ with the ones obtained within the {\rm %
HF-RPA}, $\rho _{{\rm tr}}^{{\rm HF-RPA}}({\bf r})$, and the {\rm FLA}, $%
\rho _{{\rm tr}}^{{\rm FLA}}({\bf r})$, approaches for the main resonance $%
L=0$ and its overtone. The {\rm FLA} transition density is given by Eq. (\ref
{tr4}) with the wave number $q$\ obtained from the secular equation (\ref
{sec2}). The contribution of the {\rm ISGMR} overtone to the {\rm EWSR} for
the case of the transition operator $\hat{f}_{\xi }({\bf r})$ is given by 
\begin{equation}
m_{02}(\xi )=E_{02}\left| \int d{\bf r}\ \hat{f}_{\xi }({\bf r})\rho _{{\rm %
tr,}02}^{{\rm FLA}}({\bf r})\right| ^{2},  \label{m02}
\end{equation}
where $E_{02}=\hbar \omega (q_{02})$. The eigenfrequency $\omega (q_{02})$
is obtained from the dispersion equation\ (\ref{disp}) and the wave number $%
q_{02}$ is the second (overtone) solution to the secular equation (\ref{sec2}%
). The {\rm EWSR} for the transition operator $\hat{f}_{\xi }({\bf r})$
reads 
\begin{equation}
m_{1}(\xi )=\frac{\hbar ^{2}}{2m}\int d{\bf r\ }\rho _{{\rm eq}}\left( {%
\bbox{\nabla}}\hat{f}_{\xi }({\bf r})\right) ^{2}.  \label{m102}
\end{equation}
Now, we will determine the parameter $\xi $\ from the condition that 
the value of the ratio of the partial sum $m_{02}(\xi )$ to the total sum\
$m_{1}(\xi )$ is the maximum. In {\rm  Fig. 1,} we have plotted the ratio $m_{02}(\xi
)/m_{1}(\xi )$\ as a function of the parameter $\xi $ for the nucleus
$^{208}${\rm Pb} obtained in the {\rm FLA} (dashed line) and the {\rm
HF-RPA} (solid line) models .  As seen from {\rm Fig. 1}, 
the maximum value of the ratio of\ $m_{02}(\xi )$\ to \ $m_{1}(\xi )$\
 for the FLA model is achieved for $\xi =68.3$ ${\rm fm}^{2}$ where the overtone exhausts
about $73\%$ of the {\rm EWSR}. In the case of {\rm HF-RPA}, the  maximum
ratio\ is achieved for $ \xi =78.6$ ${\rm fm}^{2}$ where the overtone,
in the energy range of 25 - 50 MeV, exhausts about $60\%$ of the {\rm EWSR}.  
It is interesting to note that if we use the FLA and HF ground-state densities
to calculate  $\xi$ from  Eq. (\ref{xi02}), we get $\xi = 
72.2$ and 79.0 fm$^2$, respectively. These values  are close to the  corresponding ones
(68.3 and 78.6 fm$^2$) obtained by the condition of  maximizing the ratio $m_{02}(\xi)/m_1(\xi)$.
 This means  that the mixing
parameter $\xi $ \ in the transition operator $  \hat{f}_{\xi }({\bf
r})$\ of Eq. (\ref{f4})\ can also be derived from the condition that
the main {\rm ISGMR} gives a minimal contribution to the energy-weighted
sum rule  $m_{1}(\xi )$.  The FLA as well as HF-RPA  calculations show that
the difference in the values of $\xi $\ obtained in both  conditions does
not exceed $\sim 0.2\%$.  We have also  calculated the dependence of
the parameter $\xi $\ on the nuclear mass number $A$.
Following the same procedure we find for $^{90}$Zr, $^{116}$Sn and $^{144}$Sm
nuclei the value of mixing parameter $\xi  = $ 38.6 (48.5), 45.9 (56.7)
and  53.1 (66.1) fm$^2$ from the FLA (RPA) calculations, respectively. These values can be
well approximated by $\xi = 1.89{\rm A}^{2/3}$ and 2.36${\rm A}^{2/3}$ fm$^2$
for the FLA and RPA approaches, respectively.

In Fig. 2, we plot the FLA and RPA results for the fraction
energy-weighted transition strength as a function of the excitation
energy obtained for the transition operator $\hat{f}_{\xi}({\bf r})$
for the $^{208}$Pb nucleus. We use $\xi=$ 68.3 and 78.6 \ {\rm fm}$^2$ in the
FLA and RPA calculations, respectively.  One can  clearly see that the RPA
calculation yields a wide resonance of width of  $\sim 10$ MeV around the
excitation energy $30-35$ MeV which corresponds to the ISGMR overtone.
Whereas, in the case of FLA, the transition operator $\hat{f}_\xi(\bf{r})$
(Eq. (\ref{f4})) with an appropriate value of $\xi$ gives  rise to a
well defined resonance for the overtone mode.
We also notice that the RPA results have the reminiscence of the ISGMR
main tone but it is practically eliminated in the FLA calculations.

In {\rm Figs. 3}$a${\rm \ }and {\rm 3}$b$,  we compare the radial macroscopic
transition density $\rho _{{\rm tr}}^{{\rm macr}}(r)$\  of Eq. (\ref
{tr9}), obtained using the HF ground-state density,  and the corresponding
FLA and HF-RPA ones for the overtone of the
ISGMR. The radial transition density $\rho _{{\rm tr}}(r)$ for a certain
multipolarity $L$\ is given by $\rho _{{\rm tr}}({\bf r})=\rho _{{\rm tr}%
}(r)Y_{L0}({\bf \hat{r}}).$ 
The macroscopic transition densities for the overtone in Figs. 3a and 3b
are not the same. Because, for an appropriate comparison, in Fig. 3a,
we have plotted $\rho_{\rm tr}^{\rm macr}(r)$ obtained using $\xi=78.6 \ {\rm
fm}^2$ and it is normalized to $60\%$ of the EWSR. On the other hand,
$\rho_{\rm tr}^{\rm macr}(r)$ plotted in Fig. 3b corresponds to  $\xi=68.3 \ {\rm
fm}^2$ and is normalized to $73\%$ of the EWSR.
Further, the RPA transition density is calculated by averaging over the 
energy range of 25 - 50 MeV. 
Notice that\ the shift of the nodes of $\rho _{%
{\rm tr,}02}^{{\rm FLA}}(r)$ to the left with respect to the ones of $\rho _{%
{\rm tr}}^{{\rm macr}}(r)$ is caused by the fact that in contrast to Eq. (%
\ref{tr9}) used for the macroscopic transition density,  in the FLA\  we use 
sharp nuclear surface, see Eq. (\ref{tr4}).
We also looked into the energy dependence of the RPA  transition density
for the operator $\hat{f}_\xi({\bf r})$ over the range of energy employed
for the averaging. We see that over the entire range considered, the
transition density has two nodal structure and the distance between
the nodes decreases with the increase in energy. For example,
averaging over $\Delta E=0.5$ MeV range, we find that  at the
excitation energies of 30, 40 and 50 MeV, the distances  between the
two nodes are 3.1, 2.7, and 2.4 fm, respectively, which reflects the fact that 
the microscopic transition density is state dependent.

We have used the microscopic transition densities for the operator
$\hat{f}_\xi({\bf r})$ to evaluate the cross-section for the ISGMR
overtone mode excited via inelastic scattering of $\alpha$-particles with
energies 240 and 400 MeV. 
We have used the folding model (FM)-DWBA to calculate
the excitation cross-section (see Ref. \cite{Kolomietz00} for details). We find
that for the $\alpha$-particles with 400 MeV energy, the calculated
cross-section is about 7 - 10 times higher than the one obtained for
$\alpha$-particles with 240 MeV. Note that for a monopole resonance
the cross-section is maximum at $0^o$. 
The values of the cross-section at $0^o$ for   the peak energy of the ISGMR overtone   are 0.5 and 3.5
mb/(sr MeV) for the case of 240 and 400 MeV, respectively. We 
point out that the maximum cross-section for the case of 240
MeV $\alpha$-particles  is below the current  experimental sensitivity of about
2 mb/(sr MeV) \cite{Youngbloodp}. It may be possible to identify the ISGMR overtone mode with 400
MeV $\alpha$-particles.

The transition density $\rho {\bf _{{\rm tr}}(r)}$ for the compression modes
is distributed over the nuclear interior and has a node close to the nuclear
surface for both the main ISGMR and its overtone. The transition density $%
\rho _{{\rm tr},02}{\bf (r)}$ of the overtone has an additional node in the  nuclear
interior. This feature of $\rho {\bf _{{\rm tr}}(r)}$ can be tested by
 evaluating  the strength distribution $S_{0}(k)$ of the electromagnetic
operator $j_{L}(kr)Y_{L0}({\hat{r}})$. The strength function $%
S_{0n}(k)$ for a certain eigenstate $n$\ is given by 
\begin{equation}
S_{0n}(k)=\left| I_{0n}(k)\right| ^{2}  \label{slk1}
\end{equation}
where
\begin{equation}
I_{0n}(k)= \int d{\bf r\ }\rho _{{\rm tr,}0n}{\bf (r)\ }%
j_{0}(kr)Y_{00}({\hat{r}}). \label{slk2}
\end{equation}
The strength function $S_{0n}$ is related to the excitation function of electron-nucleus
scattering in the Born approximation. We use Eqs. (\ref{slk1}) and (\ref{slk2})  to calculate
the energy-weighted sums $m_{01}(k)$ and $m_{02}(k)$ for the main ISGMR and
its overtone, respectively. In Fig. 4, we display the $k$ dependence of the
fraction energy-weighted sums $m_{01}(k)/m_{1}(k)$ and $m_{02}(k)/m_{1}(k)$
for the $^{208}${\rm Pb} nucleus obtained from the FLA (dashed line) and HF-RPA (solid
line) approaches. It can be seen from Fig. 4 that $m_{01}(k)/m_{1}(k)$ and $%
m_{02}(k)/m_{1}(k)$ depend strongly on $k$. A shift of the maximum of the
ratio $m_{02}(k)/m_{1}(k)$ for the overtone to the higher value of wave
number $k$ is due to the more complicated nodal structure of the transition
density associated with the overtone as  compared with the main
resonance. This shift can be exploited to separate the modes in
electron-nucleus scattering. In Fig. 5 we plot the surface and the volume
contributions to the integral in Eq. (\ref{slk2}) for the transition density
associated with overtone mode (see Eq. (\ref{tr4})). For smaller $k$, there
is a cancellation between the surface and the volume contributions leading
to a peak structure for the overtone response as shown in Fig. 4.

\bigskip

\section{Summary and conclusions}

\bigskip

Starting from the local strength function $S({\bf r},E)$
and using the smearing procedure, we have extended the quantum expression
for the transition density $\rho _{{\rm tr,n}}({\bf r})$ to the case of a
group of the thin-structure resonances which are localized in the GMR region
and are excited due to the specifically chosen transition operator $\hat{F}(\{%
{\bf r}_{i}\})=\sum_{i=1}^{A}\hat{f}({\bf r}_{i})$. Our approach was applied
to the study of the transition density of the ISGMR overtone. In this case,
an appropriate form of the transition operator $\hat{f}({\bf r})$ is given
by $\hat{f}({\bf r})=\hat{f}_{\xi }({\bf r})=(r^{4}-\xi r^{2})Y_{00}({\bf 
\hat{r}})$, see Eq. (\ref{f4}). The mixing parameter $\xi $ was determined
from the condition that the transition operator $\hat{f}_{\xi }(%
{\bf r})$\ provides for the single overtone the maximum fraction of the
energy-weighted sum rule  $m_{1}$\ of Eq. (\ref{m1}). The mixing parameter $\xi $%
\ depends on the nuclear mass number $A$. This dependence is well
approximated by $\xi \approx 2\cdot A^{2/3}\ {\rm fm}^{2}$.

We have applied our smearing procedure (using $\hat {f}_\xi({\bf r})$ 
associated with the maximum FEWSR of the  overtone) to the evaluation of
the smeared out transition density $\widetilde{\rho }_{{\rm tr,R}}({\bf r})$
of Eq. (\ref{tr2})\ within the {\rm HF-RPA}. We have
shown that the smearing procedure for the ISGMR overtone region provides a
simple two nodal structure of $\widetilde{\rho }_{{\rm tr,R}}({\bf r})$ (see
the solid line in Fig. 3$a$), as expected for the $L=0$\ overtone.
Moreover, the  transition density $\tilde{\rho}_{\rm tr,R}({\bf r})$, obtained
by the averaging over many quantum states, resembles its macroscopic
counterpart.  This fact is well illustrated in Fig. 3$a$ by comparing
 the quantum smeared transition
density $\widetilde{\rho }_{{\rm tr,R}}({\bf r})$ with the macroscopic one $%
\rho _{{\rm tr}}^{{\rm macr}}({\bf r})$ of Eq. (\ref{tr9}). An independent
derivation of the smeared out transition density $\widetilde{\rho }_{{\rm %
tr,R}}({\bf r})$ can be also obtained using the semiclassical approaches.\
In Sec. IV, we have applied a simple semiclassical Fermi-liquid
approximation to the evaluation of the smeared out (in quantum mechanical
sense) transition density $\rho _{{\rm tr}}^{{\rm FLA}}({\bf r})$. We have
used the same form of the transition operator\ $\hat{f}_{\xi }({\bf r})$ as
in the case\ of quantum {\rm HF-RPA} calculation to provide an additional
check of the derivation of the mixing parameter $\xi $ from the 
energy-weighted sum $m_{1}$. We found  a good agreement between the
values of parameter $\xi $ obtained in both the quantum and the
semiclassical approaches.
It is important to emphasize that equation (\ref{xi02})
together with Eq. (\ref{tr5}) provides a  simple expression for  the
macroscopic transition density that can be employed in the folding model-DWBA
analysis of excitation cross-section of the ISGMR overtone.

The nodal structure of the semiclassical transition density $\rho _{{\rm tr}%
}^{{\rm FLA}}({\bf r})$\ is similar to  that of both the quantum, $%
\widetilde{\rho }_{{\rm tr,R}}({\bf r})$, and the macroscopic, $\rho _{{\rm %
tr}}^{{\rm macr}}({\bf r})$, cases, see Figs. 3$a$ and 3$b$.\ A discrepancy
occurs in the surface region, where the particular behavior of $\rho _{{\rm %
tr}}^{{\rm FLA}}({\bf r})$ is due to the assumption of the  sharp surface of the nucleus 
in the FLA model. This discrepancy is not so significant in the
integral quantities like the strength functions. This is
illustrated in Fig. 4 for the case of the nuclear response to the
electromagnetic-like external field $\sim j_{0}(kr)Y_{00}({\hat{\bf r}})$. The
ratios $m_{01}(k)/m_1(k)$ and $m_{02}(k)/m_1(k)$ for the ISGMR and its overtone, respectively, 
show an distinct feature  in the $k$-dependence. Namely, for a certain
value of the wave number $k$, the strength function for the overtone
reaches a maximum whereas the contribution of the main resonance to the
strength function is strongly suppressed.\ This fact can be exploited  
 to separate the ISGMR and the overtone modes in electron-nucleus
scattering by varying   the electron's momentum transfer $k$.

\bigskip

\section{Acknowledgments}

\bigskip

This work was supported in part by the US Department of Energy under grant
\# DOE-FG03-93ER40773. One of us (V.M.K.) thank the Cyclotron
Institute at Texas A\&M University for the kind hospitality.

\newpage
\begin{center}
{\Large Figure captions}
\end{center}

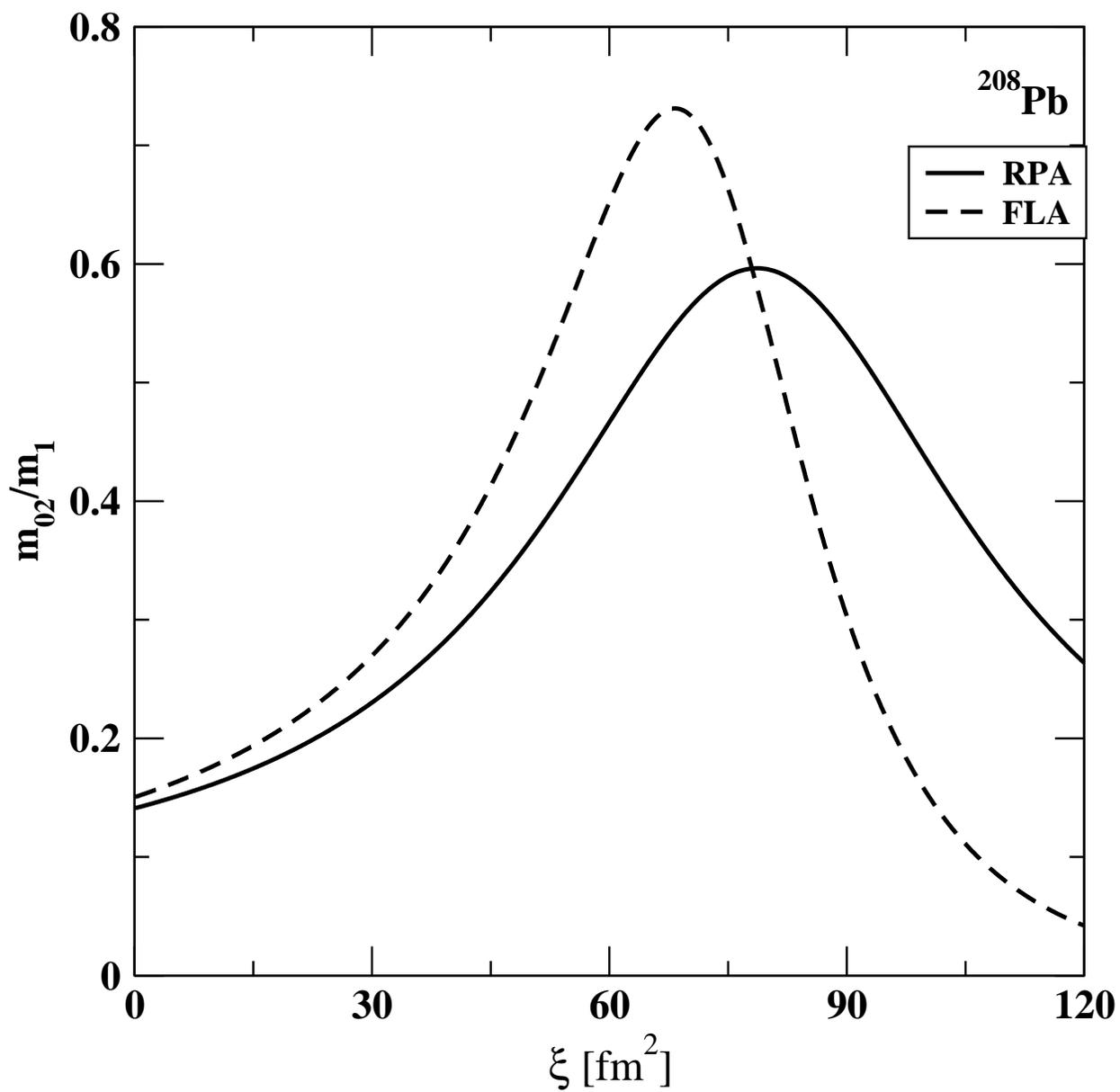
Fig. 1. The ratio $m_{02}(\xi )/m_{1}(\xi )$ of the partial contribution of the
overtone to the EWSR\ as a function of the parameter $\xi $ in the
transition operator $\hat{f}_{\xi }({\bf r})$ (Eq. (\ref{f4})) obtained within
the FLA (dashed line) and the HF-RPA (solid line) approaches,  for the monopole mode $%
L=0 $ in the nucleus $^{208}{\rm Pb.}$

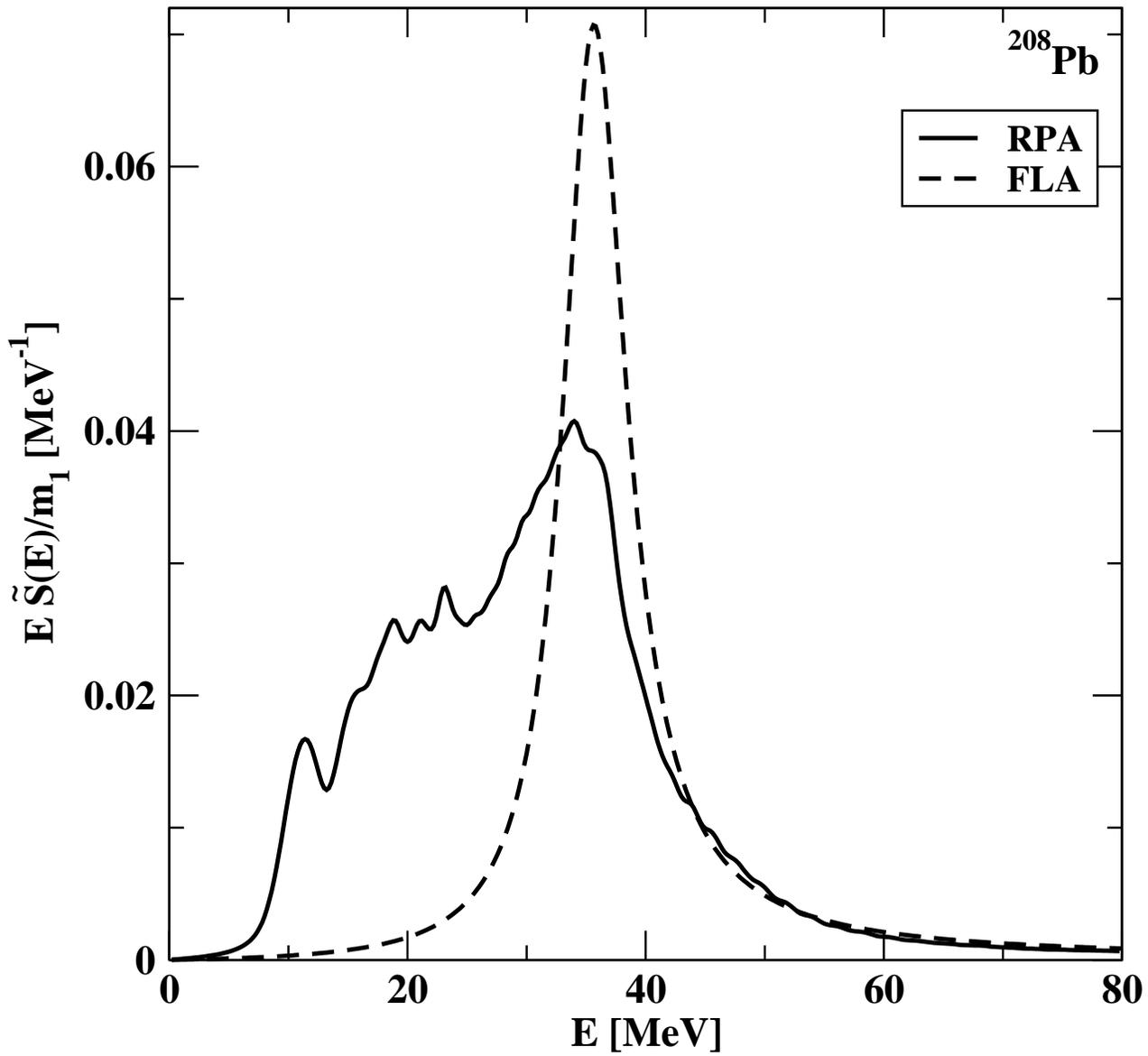
Fig. 2. The FLA and  HF-RPA results for the fraction energy-weighted transition strength for the operator 
$\hat{f}_\xi({\bf r})$ with $\xi=78.6 \ {\rm MeVfm}^2$.

Fig. 3$a$. The HF-RPA transition density, $\widetilde{\rho }_{{\rm tr}}^{%
{\rm HF-RPA}}(r),$ multiplied by $4\pi r^{2}$\ for the overtone of the ISGMR
in the nucleus $^{208}{\rm Pb}$\ (solid line) and the corresponding
macroscopic transition density $\rho _{{\rm tr}}^{{\rm macr}}({\bf r})$
taken at $\xi =78.6$ fm$^{2}$ \ (dotted line).

Fig. 3$b$. The FLA transition density, $\rho _{{\rm tr}}^{{\rm FLA}}(r)$,
multiplied by $4\pi r^{2}$\ for the overtone of the ISGMR in the nucleus $%
^{208}{\rm Pb}$\ (dashed line) and the corresponding macroscopic transition
density $\rho _{{\rm tr}}^{{\rm macr}}({\bf r})$ taken at $\xi =68.3$ fm$%
^{2}$\ (dotted line).

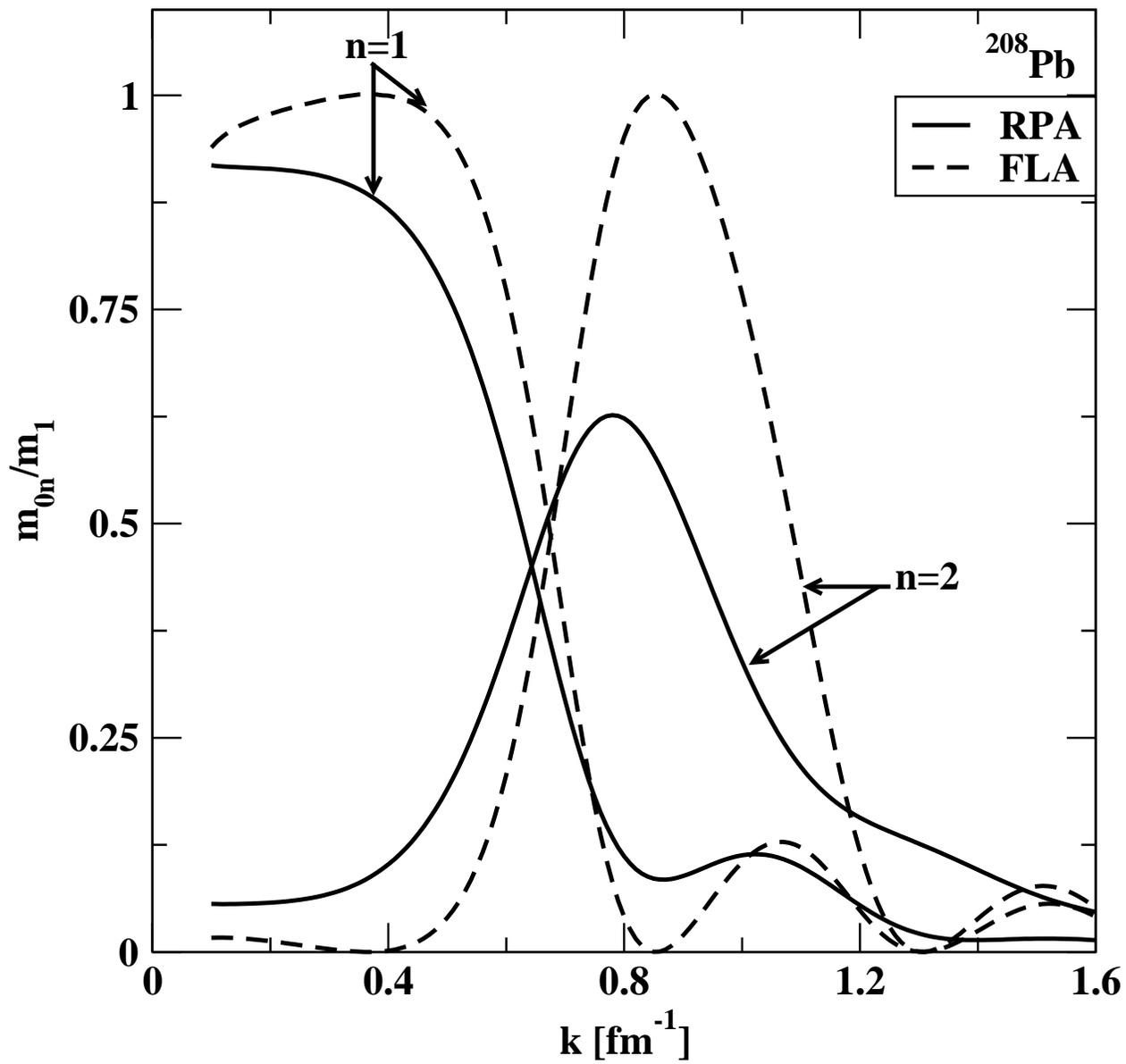
Fig. 4. The ratio $m_{0n}(k)/m_{1}(k)$ for the main ($n=1$)  ISGMR and
its overtone ($n=2$), respectively, as a function of the wave number $%
k$ obtained for electromagnetic operator $j_{0}(kr)Y_{00}(\hat{r})$ for
the $ ^{208}${\rm Pb} nucleus. The dashed and the solid lines represent
the FLA and HF-RPA results, respectively.

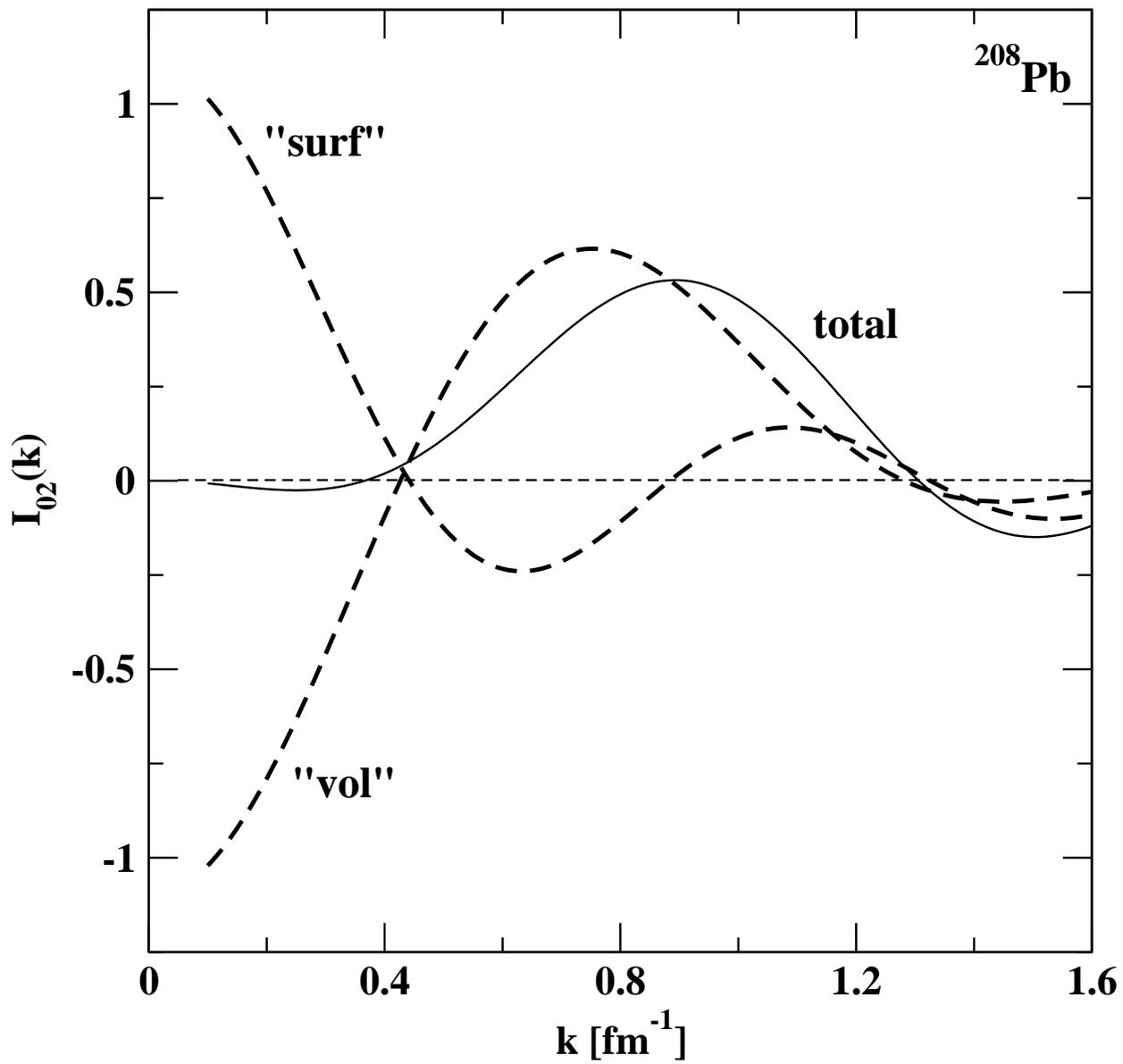
Fig. 5. Partial contributions of the volume (''vol'') and surface (''surf'')
terms of the FLA transition density of the ISGMR overtone (see Eq. (\ref{tr4}%
)) to the integral in Eq. (\ref{slk2}) (dashed lines). The solid line shows
the sum of both the volume and the surface terms.

\newpage
\begin{table}
\caption{\label{fla-rpa} 
Comparison of the FLA and RPA results for the centroid energies (in MeV)
for the main ($E_{01}$)  and the overtone ($E_{02}$) modes of ISGMR.
In the case of RPA calculations the values of $E_{01}$  are obtained by
integrating the strength function for the operator $r^2 Y_{00}$ over the
energy range of 0 - 60 MeV and  the values of $E_{02}$ are obtained using
the operator $(r^4 -\xi r^2)Y_{00}(\hat{\bf r}$) and the energy ranges
of 35 - 60, 28 - 60, 27 - 55 and 25 - 50 MeV for $^{90}Zr$, $^{116}Sn$,
$^{144}Sm$ and $^{208}Pb$ nuclei, respectively.  The experimental data
for the main tone is taken from the Ref. \cite{Youngblood99}.  }

\begin{tabular}{|c|ccc|ccc|c|}
\hline
\multicolumn{1}{|c|}{}&
\multicolumn{3}{|c|}{FLA}&
\multicolumn{3}{|c|}{RPA}&
\multicolumn{1}{|c|}{EXP.}\\
\cline{2-8}
Nucleus & $E_{01}$&$ E_{02}$ & $E_{02}/E_{01}$ &  $E_{01}$& $E_{02}$ & $E_{02}/E_{01}$ & $E_{01}$\\
\hline
$^{90}$Zr & 19.6& 43.0& 2.2& 18.1& 43.8& 2.4& 17.89$\pm$0.20\\
$^{116}$Sn & 18.2& 40.4& 2.2& 16.5& 39.1& 2.4& 16.07$\pm$0.12\\
$^{144}$Sm & 17.0& 38.5& 2.3& 15.7& 36.8& 2.3& 15.39$\pm$0.28\\
$^{208}$Pb & 15.3& 35.5& 2.3& 13.8& 33.7& 2.4& 14.17$\pm$0.28\\
\hline
\end{tabular}
\end{table}

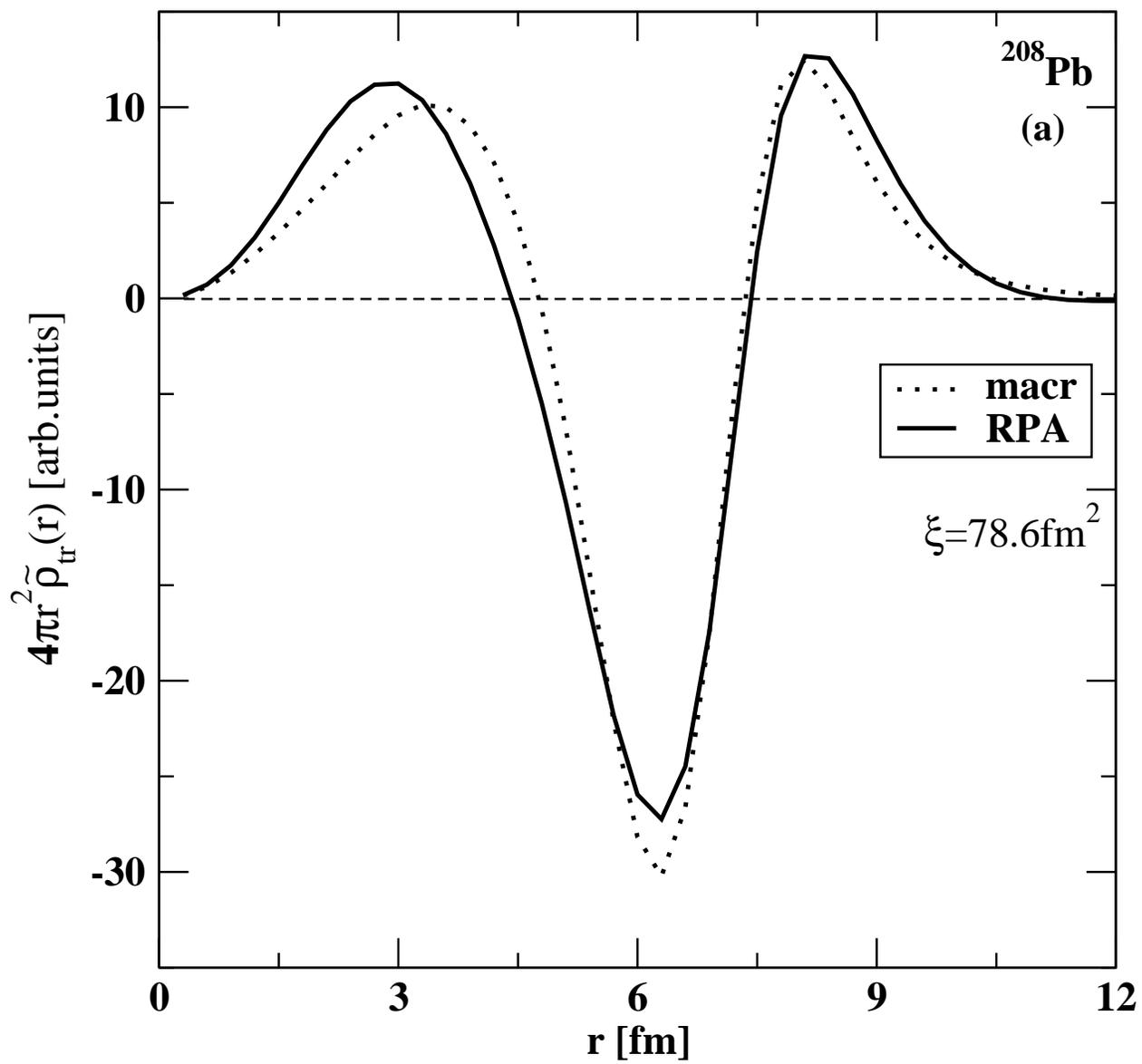
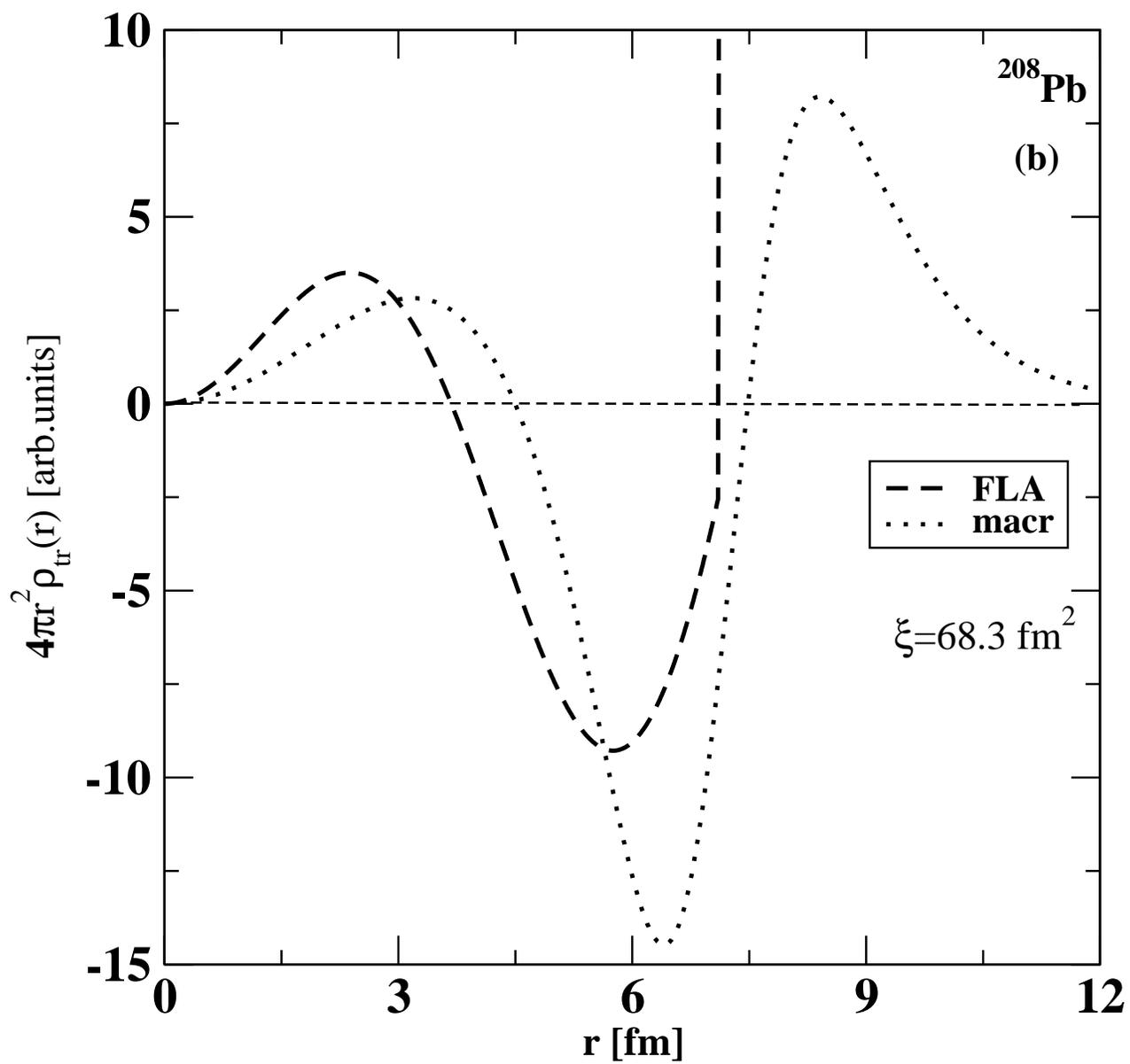
\end{document}